\documentclass[manuscript]{aastex61}
\usepackage{here}
\usepackage{booktabs}
\usepackage{amsmath,amssymb,amsfonts}
\usepackage{times}
\def\vector#1{\mbox{\boldmath $#1$}}  

\shorttitle{Chromospheric magnetic field and NLFFF}
\shortauthors{Kawabata et al.}

\begin{document}
\title{Chromospheric magnetic field: A comparison of He I 10830 \AA \  observations with nonlinear force-free field extrapolation}
\author{Yusuke Kawabata}
\affiliation{National Astronomical Observatory of Japan, 2-21-1 Osawa, Mitaka, Tokyo 181-8588, Japan}

\author{Andr\'es Asensio Ramos}
\affiliation{Instituto de Astrof\'isica de Canarias, C/V\'ia L\'actea s/n, 38205, La Laguna, Tenerife, Spain}
\affiliation{Departamento de Astrof\'isica, Universidad de La Laguna, 38206, La Laguna, Tenerife, Spain}
\author{Satoshi Inoue}
\affiliation{Institute for Space-Earth Environmental Research, Nagoya University, Furo-cho, Chikusa-ku, Nagoya, 464-8601, Japan}

\author{Toshifumi Shimizu}
\affiliation{Institute of Space and Astronautical Science, Japan Aerospace Exploration Agency, 3-1-1 Yoshinodai, Chuo, Sagamihara, Kanagawa 252-5210, Japan}
\affiliation{Department of Earth and Planetary Science, The University of Tokyo, 7-3-1 Hongo, Bunkyo-ku, Tokyo 113-0033, Japan}
\email{kawabata.yusuke@nao.ac.jp}


\begin{abstract}
The nonlinear force-free field (NLFFF) modeling has been extensively used to
infer the three-dimensional (3D) magnetic field in the solar corona. One of the
assumptions in the NLFFF extrapolation is that the plasma beta is low, but this
condition is considered to be incorrect in the photosphere. We examine 
direct measurements of the chromospheric magnetic field in two active regions
through spectropolarimetric observations at He {\sc i} 10830 \AA, which are compared with the potential fields and NLFFFs extrapolated from the photosphere. 
The comparisons allow quantitative estimation of the uncertainty
in the NLFFF extrapolation from the photosphere. Our analysis shows that
observed chromospheric magnetic field may have larger non-potentiality compared to the
photospheric magnetic field. Moreover, the large non-potentiality in the
chromospheric height may not be reproduced by the NLFFF extrapolation from the
photospheric magnetic field. The magnitude of the underestimation of the
non-potentiality at chromospheric heights may reach 30-40 degree in 
shear signed angle in some locations. 
 This deviation may be caused by the non-force-freeness in the photosphere. 
Our study suggests the importance of the inclusion of measured chromospheric magnetic fields in the NLFFF modeling for the improvement of the coronal extrapolation.
\end{abstract}



\section{Introduction}
Many energetic events, such as solar flares and coronal mass ejections (CMEs),
are caused by the release of the magnetic field energy stored in the solar corona.
 Obtaining the magnetic field information in the solar atmosphere is crucial task to understand the mechanisms of such dynamic events.
The solar magnetic field is routinely measured via spectropolarimetric observations in the photosphere and chromosphere \citep{2017SSRv..210..109D}.
 On the other hand, the coronal magnetic field is difficult to obtain due to the lower brightness of the coronal lines.

The force-free field modeling is one of the alternative methods to infer the three-dimensional (3D)
magnetic field in the solar corona. The main concept of the force-free field
modeling is to extrapolate the magnetic field lines from the spatial map of
the magnetic field in the photosphere  based on two assumptions; low plasma beta
and mechanical equilibrium in the solar corona  \citep{2012LRSP....9....5W}. The
two assumptions lead to the condition that the Lorentz force vanishes in the
solar corona, i.e., the magnetic tension and the magnetic pressure are balanced.
That is,
\begin{equation}
\vector{j}\times\vector{B}=0,
\label{fff}
\end{equation}
where $\vector{j}$ is the current density, and $\vector{B}$ is the magnetic
field. Eqn. (\ref{fff}) can be rewritten by using Amp\'ere's law
$\nabla\times\vector{B}=4\pi \vector{j}/c$, so that
\begin{equation}
\nabla\times\vector{B}=\alpha\vector{B},
\end{equation}
where $\alpha$ is the force-free parameter. When $\alpha$ has a spatial
dependence, the magnetic field distribution is known as \emph{nonlinear force-free field}
(NLFFF).

One of the most controversial problems in NLFFF extrapolation is the assumption of force-freeness in the photosphere.
From the model of \cite{2001SoPh..203...71G}, the plasma beta in plage
regions in the photosphere is of order of $10^2$, while the magnetic field at the
center of sunspots is
almost force-free. There are some previous studies to investigate the force-freeness
in active regions in the photosphere based on the 
force-free condition derived by \cite{1985svmf.nasa...49L}. The Lorentz force
can be written as the divergence of the Maxwell stress tensor, 
\begin{equation}
M_{ij}=-\frac{B^2}{8\pi} \delta_{ij}+\frac{B_{i}B_{j}}{4\pi}.
\end{equation}
Assuming that the strength of magnetic field vanishes at very large heights, three cartesian components of the volume-integrated Lorentz force can be approximately 
written with the following surface integrals:
\begin{eqnarray}
F_x&=&\frac{1}{4\pi}\int B_{x}B_{z} \textrm{d}x \textrm{d}y , \label{fx}\\
F_y&=&\frac{1}{4\pi}\int B_{y}B_{z} \textrm{d}x \textrm{d}y ,  \label{fy}\\
F_z&=&\frac{1}{8\pi}\int (B_{z}^2-B_{x}^2-B_{y}^2) \textrm{d}x \textrm{d}y.  \label{fz}
\end{eqnarray}
According to \cite{1985svmf.nasa...49L}, the magnetic field is
force-free if the three components of the net Lorentz force are 
smaller than the integrated magnetic pressure force:
\begin{equation}
F_{p}=\frac{1}{8\pi}\int (B_{x}^2+B_{y}^2+B_{z}^2) \textrm{d}x \textrm{d}y. \label{fp}
\end{equation}

\cite{1995ApJ...439..474M} investigated the force-freeness in the photosphere
and the chromosphere using observations of the Na \textsc{I} 5896 \AA\ line. They showed that while
$|F_{z}|/F_p\sim0.4$ in the photosphere,  $|F_{z}|/F_p$ becomes 0.1 roughly
400~km above the photosphere and concluded that the photosphere is not force-free
while the chromosphere is indeed force-free. \cite{2002ApJ...568..422M} analyzed 12 magnetograms
obtained from the Fe {\sc i} pair of lines at 6301.5 and 6302.5 \AA\ and showed that the value of
$|F_{z}|/F_p$ ranges from 0.06 to 0.32 with a median value of 0.13. This result
implies that the photospheric magnetic field is not far from the force-free
state. On the other hand, \cite{2013PASA...30....5L} performed a statistical study
of the force-freeness using 925 magnetograms and found that only 25\% of the active
regions satisfy $|F_{z}|/F_p < 0.1$. 
We note that the conditions described by \cite{1985svmf.nasa...49L}
are not sufficient for force-freeness.

The validity of the NLFFF modeling has been checked by X-ray and/or extreme ultraviolet (EUV)
imaging observations. Imaging observations have the disadvantage that
quantitative information about the magnetic field strength can not be obtained.
Moreover, X-ray and/or EUV observations suffer from projection effects and
are sensitive to the presence of multiple loops along the line of sight. Therefore, how the non-force-freeness in the
photosphere can affect the 3D configuration of the magnetic field in the NLFFF
modeling is still unclear.

To reveal the NLFFF uncertainty in the upper atmosphere, we make use of spectropolarimetric observations with chromospheric spectral lines for two active regions and derive the chromospheric magnetic field. 
The derived chromospheric magnetic field is compared with the potential as well as NLFFF extrapolations from the photospheric magnetic field. 
Using the magnetic field in the chromosphere, being a layer between the photosphere and the corona, can help
better understand the phenomena occurring in active regions. 
First, the magnetic field in the chromosphere will play an important role in improving extrapolation methods. Although the force-freeness of the photospheric magnetic field in active regions is
controversial, the chromospheric magnetic field is thought to be
sufficiently force-free \citep{1995ApJ...439..474M, 2001SoPh..203...71G}.
Therefore, using the chromospheric magnetic field as the bottom boundary can improve
the NLFFF modeling of the coronal magnetic field. Second, we can quantitatively
compare the NLFFF modeling from both the photosphere and chromosphere and understand the
effect of the non-force-freeness. 
Third, the chromospheric magnetic field measurements will help us to understand the onset mechanisms of solar flares.
Recent works suggest that a magnetic reconnection in chromospheric layers is a suitable mechanism for the onset of solar flares \citep{2012ApJ...760...31K, 2013ApJ...778...48B, 2017NatAs...1E..85W}.

Accurate measurements of the chromospheric magnetic field are challenging. 
The magnetic field in the chromosphere has been qualitatively and quantitatively measured with ground-based telescopes using the Ca {\sc ii} H \& K lines (3934 and 3968 \AA), H$\alpha$ (6563 \AA), Ca {\sc ii}
infrared lines (8949, 8542, and 8662 \AA), and He {\sc i} lines at 5876 and 10830
\AA\ \citep[see the review of][]{2017SSRv..210..109D}.
In the solar atmosphere, the Zeeman effect and the Hanle effect are the
primary mechanisms to produce polarimetric signals in these spectral lines in
the presence of the magnetic field. The Hanle effect modifies polarization
signals which are produced by the scattering polarization when the magnetic field is
inclined with respect to the symmetry axis of the radiation field
\citep[e.g.,][]{2001ASPC..236..161T}. Compared to the Zeeman effect, the Hanle effect is
sensitive to weaker fields, typically in the range between 1 and 100 G for selected solar
spectral lines. One of the difficulties to infer the magnetic field in the
chromosphere from spectropolarimetric observations is the necessity of
the complex atmospheric model. For example, the Ca {\sc ii} K and H$\alpha$ lines
need to be modeled in non-local thermodynamic equilibrium (NLTE)
\citep{1981ApJS...45..635V}. On the contrary, He {\sc i} 10830 \AA\, which results from the
transition between the terms $2s^3S$ and $2p^3P$ of the triplet system of He \textsc{i} is
simpler to model. The
only feasible way to populate the lower term of the multiplet is via EUV radiation from the corona.
Therefore, the formation layer of He {\sc i} 10830 \AA \ is thinner compared to
other chromospheric lines, which makes it possible to interpret the line with
a simple constant slab model.
 
\cite{2003Natur.425..692S} determined the chromospheric vector magnetic field
in small emerging active regions through the inversion of the
spectropolarimetric data at He {\sc i} 10830 \AA. They revealed the existence of
a tangential discontinuity of the magnetic field direction, which is the
observational signature of an electric current sheet. The magnetic field vector
at other chromospheric features has also been studied; active region
filaments \citep{2012ApJ...749..138X}, superpenumbral fine structures
\citep{2013ApJ...768..111S,2015SoPh..290.1607S}, and sunspots
\citep{2017A&A...604A..98J}. 
 The usage of force-free magnetic field boundary conditions for the NLFFF modeling has also been investigated. 
One of the methods is the preprocessing method, with which we can obtain magnetic fields similar to those in the chromosphere from photospheric observations.
This preprocessing method was firstly proposed by \cite{2006SoPh..233..215W}, which
consists of minimizing the total force and torque on the bottom boundary. 
\cite{2012ApJ...752..126Y} developed a new preprocessing method with chromospheric magnetic field.  
They improved the method by adding a new term concerning chromospheric longitudinal fields. 
They found that some preprocessed fields display the
smallest force- and torque-freeness. \cite{2012ApJ...748...23Y} investigated the three-dimensional structure of an active region filament by using
NLFFF extrapolations based on simultaneous observations at photospheric (Si
{\sc i} 10827 \AA) and chromospheric (He {\sc i} 10830 \AA) heights. The
extrapolations yield  a twisted flux rope whose axis is
located at about 1.4 Mm above the solar surface.

In this work we investigate how the magnetic field is distributed at chromospheric heights
and check the reliability of the NLFFF modeling. Although previous
studies revealed some properties of the chromospheric magnetic field, the field-of-views (FOVs)
of their observations were limited because the seeing made it difficult to
perform the stable large FOV scanning. This turns out to be important for a 
reliable comparison between the NLFFF extrapolation and the chromospheric magnetic field. 
By analyzing the chromospheric magnetic field in whole
active regions, we attempt to reveal the non-potential magnetic field
distribution in the chromosphere through spectropolarimetric observations at He
{\sc i} 10830 \AA \ and how significantly the magnetic field at the
chromospheric height derived by the current NLFFF modeling with photospheric
magnetic field is deviated from the measured chromospheric magnetic field.

The paper is organized as follows: The observations are presented in Section 2,  the data
reduction  and the method of NLFFF are described in Section 3. Section 4 presents the results, followed
by discussion in Section 5 and the summary in Section 6.


\section{Observations}
 
 \subsection{Observations of NOAA 10969}
\begin{figure}
\includegraphics[width=\columnwidth,clip,bb=0 0 1024 871]{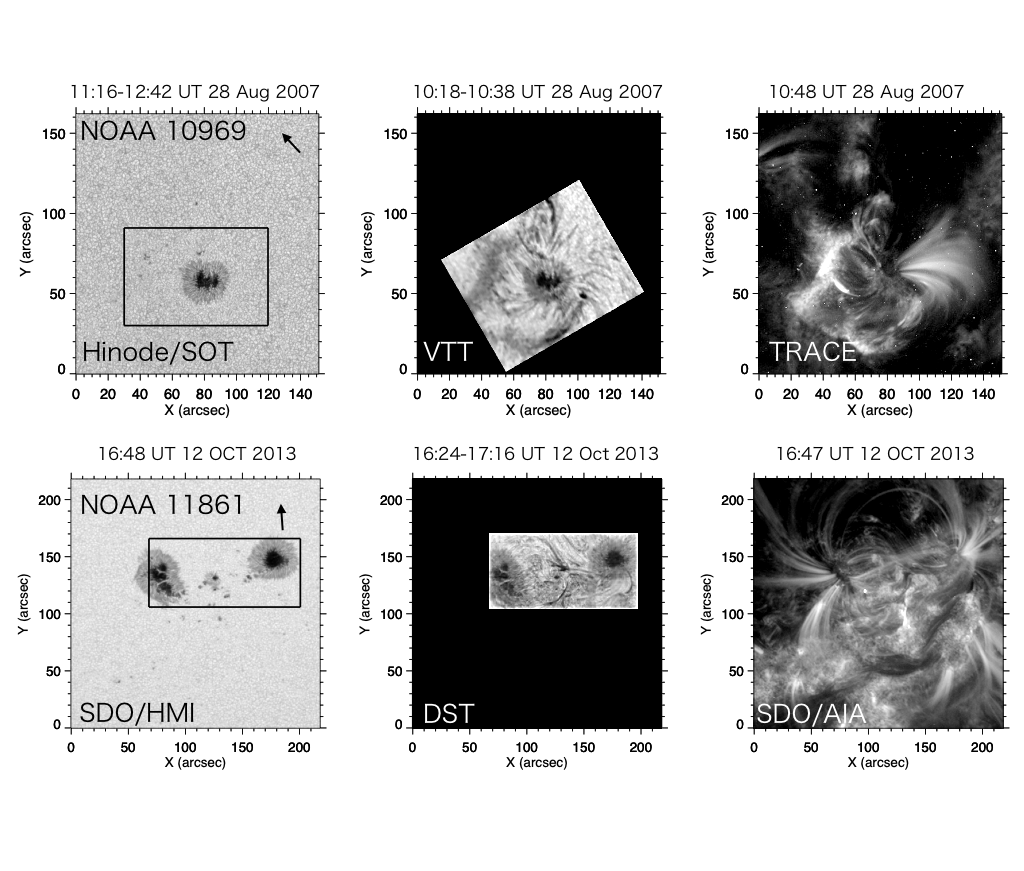}
\caption{Upper left: The continuum image obtained with {\it Hinode}/SOT SP. The black arrow shows the direction of the disk center. The black box shows the region of interest (ROI) in Figures \ref{sec3:observation_stokes_noaa10969}, \ref{sec3:bxbybz_noaa10969}, and \ref{sec3:ssa_noaa10969}. Upper middle: The line core image of  He {\sc i} 10830 \AA \ obtained with VTT/TIP-2. Upper right: The EUV image at 171 \AA \ obtained with {\it TRACE} . Lower left: The continuum image obtained with {\it SDO}/HMI. The black arrow shows the direction of the disk center. The black box shows the ROI in Figures \ref{sec3:observation_stokes_noaa11861}, \ref{sec3:bxbybz_noaa11861}, and \ref{sec3:ssa_noaa11861}. Lower middle:The line core image of  He {\sc i} 10830 \AA \ obtained with DST/FIRS. Lower right: The EUV image at 171 \AA \ obtained with {\it SDO}/AIA.}
\label{sec3:observation}
\end{figure}

NOAA active region 10969 was a simple bipolar active region as shown in the upper panels of Figure \ref{sec3:observation}.
The leading sunspot has a negative polarity and there are several positive magnetic  islands to the east of the sunspot.
 The Solar Optical Telescope \citep[SOT;][]{2008SoPh..249..167T,2008SoPh..249..221S,2008SoPh..249..197S,2008SoPh..249..233I}/Spectropolarimeter \citep[SP;][]{2013SoPh..283..579L}  aboard the {\it Hinode} satellite  \citep{2007SoPh..243....3K} measured the full Stokes vector of Fe {\sc I} 6301.5 \AA \ and 6302.5 \AA \ lines in the period between 11:16 UT and 12:42 UT on 28 Aug 2007.
 The spectral sampling is 21.5 m\AA \ per pixel.
 NOAA 10969 was located close to the disk center, i.e., (111 \arcsec, -184\arcsec) in the heliocentric coordinate at that time. 
 The map has an effective pixel size of 0$\arcsec$.16 along slit and 0$\arcsec$.15 slit step with the FOV of 152$\arcsec \times$164$\arcsec$. 
 The Stokes profiles are obtained with 6 rotation cycles (4.8 seconds) of polarization modulator unit.

NOAA 10969 was also observed by the Tenerife Infrared Polarimeter-2
\cite[TIP-2;][]{2007ASPC..368..611C} mounted on the German Vacuum Tower
Telescope (VTT) at the Observatorio del Teide, Tenerife, Spain, between 10:18-10:38
UT on 28 Aug 2007. The VTT/TIP-2 measured the full Stokes vector around the He {\sc i}
triplet at 10830 \AA \ with a spectral sampling of 11 m\AA \ per pixel. 
The exposure time was 0.25 seconds and four accumulations per modulation step were
performed. The noise level measured in continuum wavelengths was $3\times10^{-3}$ in units
of the continuum intensity. 
The active region was scanned with a 0.\arcsec18 along the slit and steps of 0.\arcsec5. 
We have to note that there might be small magnetic flux which can not be detected in this scanning because the scanning was sparse raster.

For the coronal structure, we observed the region with a spatial resolution of 1\arcsec (0\arcsec.5 pixel$^{-1}$) at 171
\AA\ obtained with the {\it Transition Region and Coronal Explorer} \citep[{\it
TRACE};][]{1999SoPh..187..229H}, which is sensitive to coronal plasma at
a temperature around 1 MK.

\subsection{Observations of NOAA 11861}

NOAA active region 11861 had multiple sunspots. The continuum image observed
with {\it SDO}/HMI at 16:48 UT on 12 Oct 2013 is shown in the bottom panel of
Figure \ref{sec3:observation}. NOAA 10969 was located close to the disk center,
i.e., (0 \arcsec, -250\arcsec) in the heliocentric coordinate at that time.

 The full Stokes vector of He {\sc i} 10830 \AA \ was obtained by the Facility Infrared Spectropolarimeter \citep[FIRS;][]{2010MmSAI..81..763J} at the Dunn Solar Telescope (DST) located on Sacramento Peak in New Mexico, USA. 
 The FIRS scanned the active region between 16:24 and 17:16 UT on 12 Oct 2013 with a spectral sampling of 39 m\AA \ per pixel. 
 The active region was scanned with 0.\arcsec15 along the slit and
 steps of 0.\arcsec3, for a total FOV of 132\arcsec$\times$66\arcsec. The exposure time
 was 0.125 seconds and four accumulations per modulation step were performed.
 The standard deviation of the continuum intensity was $1\times10^{-2}$. 
 Because the continuum contains large fringes, the standard deviation of the continuum becomes large.
 We used images with 0\arcsec.5 pixel$^{-1}$ in the 171 \AA\ channel of 
 the Atmospheric Imaging Assembly \citep[AIA;][] {2012SoPh..275...17L} onboard 
 {\it SDO} as context images for the corona.


\section{Method}

 \subsection{Data Reduction}
 
For the calibration of the {\it Hinode}/SOT SP data, we used the Solarsoft routine SP\_PREP  \citep{2013SoPh..283..601L}.
After the calibration of the spectropolarimetric data, we applied a Milne-Edington inversion by the code based on MELANIE \citep{2001ASPC..236..487S}.
The 180$^\circ$ ambiguity in the transverse magnetic field direction was solved
with the minimum energy ambiguity resolution method
\citep{1994SoPh..155..235M,2009ASPC..415..365L}. For the HMI data, we used the
vector magnetic field data product, SHARP \citep{2014SoPh..289.3549B}. For the
VTT data, flat field, dark current corrections, and the standard polarimetric
calibration were carried out \citep{1999AGAb...15Q..11C,2003SPIE.4843...55C}.
The wavelength calibration was also performed by fitting the observed spectrum with
the solar spectrum atlas \citep{1981atlasinfra}. In order to improve the
signal-to-noise ratio, we carried out a binning of 4 pixels in the spectral direction
and 4 pixels along the slit direction. The resulting noise levels of stokes $Q/I$, $U/I$,
and $V/I$ are, in units of the continuum intensity, $5.3\times 10^{-4}$, $6.3\times10^{-4}$ and $7.8\times 10^{-4}$,
respectively. For the DST data, we carried out the basic data reduction
including flat fielding, dark current corrections, polarimetric calibration and wavelength
calibration \citep{2005A&A...437.1159B}. 
Because of the presence of significant polarized fringes in the DST data, we
removed them using the pattern-recognition method of \cite{2012ApJ...756..194C}. 
A binning of 2 pixels along the spectral direction
and along the slit, and 4 pixels along the scanning direction were
carried out. The resulting noise levels of stokes $Q/I$, $U/I$, and $V/I$ are
$1.0\times 10^{-4}$, $7.2\times10^{-5}$ and $5.4\times 10^{-4}$, respectively.
An initial incomplete ambiguity resolution is carried out for the
chromospheric data assuming that only the 180$^\circ$ ambiguity is
present. This is done by choosing the azimuth closer to the potential field extrapolation.
We defer a more in-depth discussion about ambiguities of the inversion results to Section
\ref{sec3:discussion}.

The inversion of the He {\sc i} 10830 \AA\ multiplet was performed by HAZEL
\citep{2008ApJ...683..542A}, which considers the joint action of the Hanle and
Zeeman effects in a simple slab model with all physical properties constant.
Eight parameters describe such slab model of HAZEL;
the magnetic field strength, the inclination and azimuth of the magnetic field
vector, the optical depth of the slab (measured in the core of the red component), 
the height above the solar surface at which the slab lies, the Doppler width of the line,
the Doppler velocity, and the line damping parameter.
In order to reduce the computing time and also to reduce the ambiguity in the solution
space, we fix two of these parameters.
The first one is the damping parameter, which we fix to zero.
The broadening of the He {\sc i} 10830 \AA \ multiplet is dominated by Doppler broadening and the information
to fix the effect of the damping parameter lies in the far wings, which are affected by blends. 
In addition,  \cite{2004A&A...414.1109L} reported that the inclusion of the damping parameter barely affects the
inferred parameters, although it slightly improves the fit.
The second parameter is the height of the slab, which we fix to 2 \arcsec $\sim 1500$ km in this study.
The scattering polarization and Hanle effect depend on the anisotropy of the radiation field.
Because the height of the slab increases the anisotropy, the linear polarization signal is slightly 
affected by the height of the slab \citep{2011A&A...532A..63M}. However, we find no important change in the
inferred parameters even if we modify the height in sensible ranges.

 \subsection{NLFFF Extrapolation}
 NLFFF calculations were performed by the MHD relaxation method \citep{2014ApJ...780..101I}, which solves the
 following set of equations:
 \begin{eqnarray}
\frac{\partial \vector{v}}{\partial t}&=&-(\vector{v}\cdot \vector{\nabla})\vector{v}+\frac{1}{\rho}\vector{j}\times\vector{B}+\nu\nabla^2\vector{v}, \label{momentum} \\
\frac{\partial \vector{B}}{\partial t}&=&\nabla \times (\vector{v}\times \vector{B}-\eta \vector{j})-\nabla \phi, \label{induction}\\
\vector{j}&=& \nabla \times \vector{B}, \label{ampere}\\
\frac{\partial \phi}{\partial t} &+&c_{h}^2\nabla\cdot \vector{B}=-\frac{c_{h}^2}{c_{p}^2}\phi. \label{9wave}
\end{eqnarray}
Equations (\ref{momentum}), (\ref{induction}) and (\ref{ampere}) are the equation of motion,
the induction equation, and the ${\rm Amp\grave{e}re's}$ law, respectively. 
Eq. (\ref{9wave}) follows the procedure developed by \cite{2002JCoPh.175..645D} to force $\nabla\cdot\vector{B}=0$, making
use of the potential $\phi$. $\rho$ is the pseudo density, which is assumed to be equal to
$|\vector{B}|$ to ease the relaxation by equalizing the Alfv\'en speed in space.
 We chose the non-dimensional viscosity $\nu=10^{-3}$, which corresponds to $ 6\times10^5 \ {\rm m^2s^{-1}}$ for NOAA 10969 and $8\times10^5 \ {\rm m^2s^{-1}}$ for NOAA 11861  in real units. 
The length and magnetic field were normalized by $L_{0}=110$ Mm (NOAA 10969)
and $L_{0}=157$ Mm (NOAA 11861)  and $B_{0}=3000$ G, respectively.
The numerical domain for the NLFFF extrapolation is set to $(0,0,0) <(x,y,z)<(1.0,1.07,0.25)$ resolved by $504\times540\times504$ cells for NOAA10969 and $(0,0,0) <(x,y,z)<(1.0,1.0,0.75)$ resolved by
$432\times432\times648$ cells for NOAA 11861. 
$x$ and $y$ are the horizontal directions and $z$ is the vertical direction. 
The number of steps in the calculation were set to 40000 for both NOAA 10969 and NOAA 11861. 
The  velocities are normalized to the Alfv\'en velocity $V_{\rm A} \equiv B_0/(4\pi \rho_0)^{1/2}$, and times to the Alfv\'en time $\tau_{A} \equiv L_{0}/V_{A}$.  
The density $\rho_0$ in the Alfv\'en velocity is set equal to $B_0$.
The parameters $c_{p}^2=0.1$ and $c_{h}^2=0.04$ are the non-dimensional advection and
non-dimensional diffusion coefficients, which are assumed constant. The
non-dimensional resistivity $\eta$ is given by
\begin{equation}
\eta=\eta_{0}+\eta_{1}\frac{|\vector{j}\times\vector{B}||\vector{v}|^2}{|\vector{B}|^2},
\label{resistivity}
\end{equation}
where $\eta_{0}$ and $\eta_{1}$ are fixed at $5.0\times 10^{-5}$ and $1.0\times10^{-3}$ in non-dimensional units, respectively.  
 In real units, these values correspond to $3\times10^4 \ {\rm m^2s^{-1}}$ and $6\times10^5 \ {\rm m^2s^{-1}}$ for NOAA 10969, and $4\times10^4 \ {\rm m^2s^{-1}}$ and $8\times10^5 \ {\rm m^2s^{-1}}$ for NOAA 11861, respectively.
The second term is introduced to accelerate the relaxation to the force-free state.

The bottom boundary is the photospheric magnetic field observed with {\it
Hinode}/SOT SP for NOAA 10969 and {\it SDO}/HMI for NOAA 11861. The potential
field is used as an initial guess for both regions.

\subsection{Shear Signed Angle}
To evaluate the non-potentiality at each height, we use the shear signed angle (SSA), which is defined as
\begin{equation}
{\rm SSA}=\tan^{-1}\left(\frac{B_yB_{xp}-B_{yp}B_x}{B_xB_{xp}+B_yB_{yp}}\right) .
\end{equation}
The SSA is the deviation of azimuth angle from the potential magnetic field ($B_{xp}$ and $B_{yp}$).
For the chromospheric magnetic field derived from He {\sc i} 10830 \AA, the SSA is calculated by the 
potential field calculated from the $B_z$ at 1500 km height.

\section{Results}

\subsection{Polarimetric signals of He {\sc i} 10830 \AA}

\begin{figure}
\begin{center}
\includegraphics[width=0.65\columnwidth,clip,bb=0 0 566 1133]{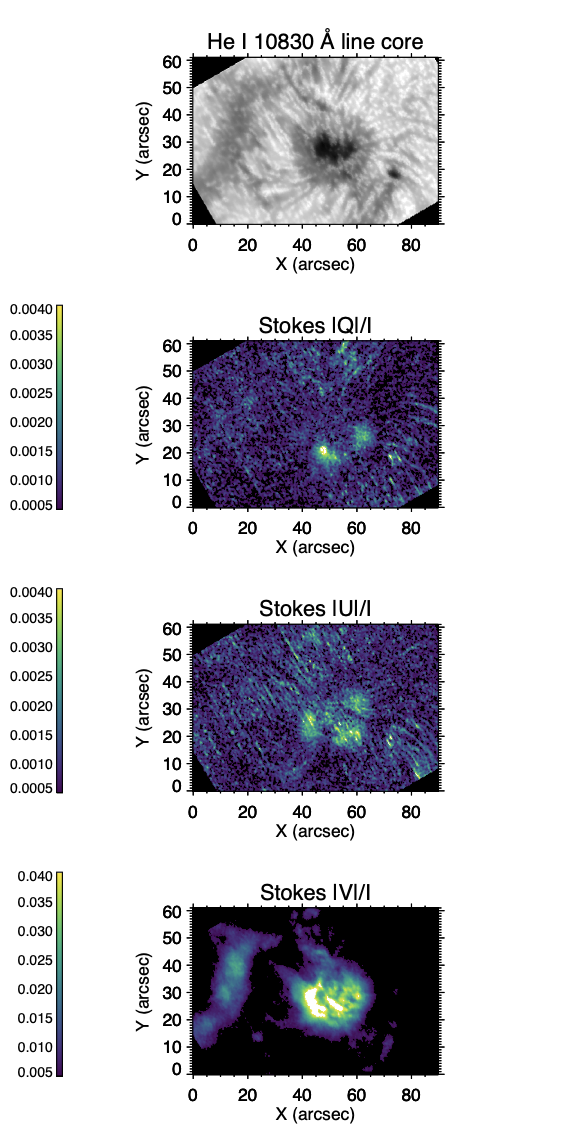}
\end{center}
\caption{The absolute peak value of Stokes Q/I, U/I, and V/I of NOAA 10969 at He {\sc i} 10830 \AA \ 
obtained with VTT/TIP-2 between 10:18-10:38 UT on 28 Aug 2007.}
\label{sec3:observation_stokes_noaa10969}
\end{figure}

Figure \ref{sec3:observation_stokes_noaa10969} shows the absolute peak values of
Stokes $Q/I$, $U/I$, and $V/I$ in NOAA 10969. The signal of Stokes $V/I$ is strong in the
leading sunspot and the magnetic islands of the positive polarities where the
vertical magnetic field exists. The strong linear polarization signals can be identified in
the outer part of the spot and the fibril structure between the positive and
negative polarities, which come from the Zeeman effect and/or the Hanle effect.

\begin{figure}
\includegraphics[width=0.6\columnwidth,clip,bb=0 0 566 1133]{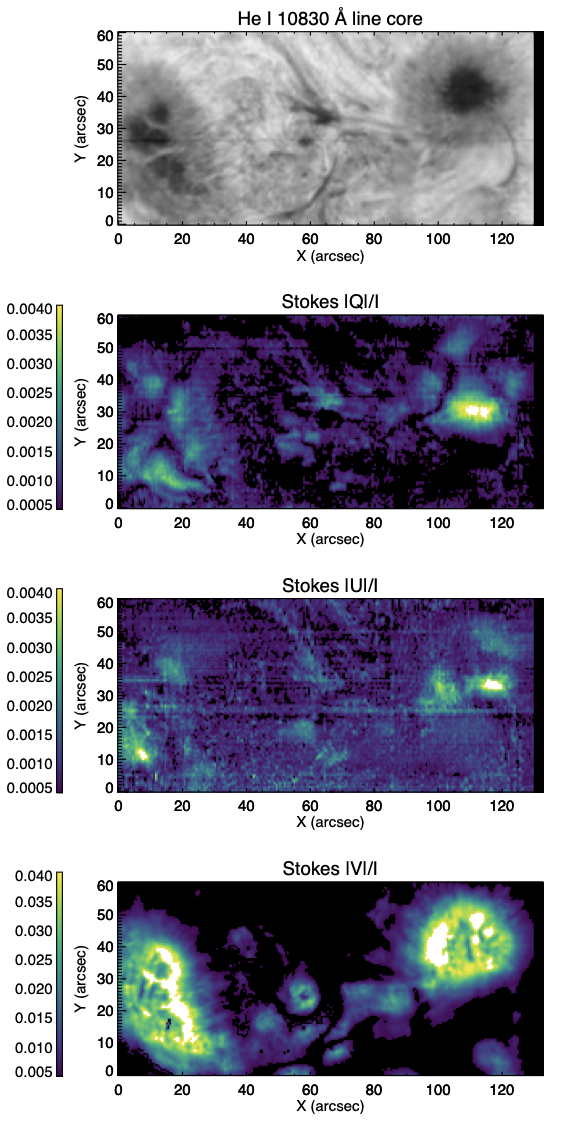}
\caption{The absolute peak value of Stokes Q/I, U/I, and V/I of NOAA 11861 at He {\sc i} 10830 \AA \ obtained with DST/FIRS between 16:24 and 17:16 UT on 12 Oct 2013. }
\label{sec3:observation_stokes_noaa11861}
\end{figure}

Figure \ref{sec3:observation_stokes_noaa11861} shows the absolute peak value of Stokes $Q/I$, $U/I$, and $V/I$ in NOAA 11861.
Strong circular polarization signals can be seen in the two large spots and small spot between them.
There are strong linear polarization signals in the penumbral regions.

\subsection{Force-Freeness of the Active Regions}
Table \ref{sec3:table:forcefreeness} shows the force-freeness based on the
Eqs. (\ref{fx}), (\ref{fy}), and (\ref{fz}) for NOAA 10969 and 11861 at photospheric height. 
As shown later, there are some pixels where the magnetic field is not derived with sufficient accuracy at the chromospheric height. 
Therefore, we do not show the force-freeness in the chromosphere in this paper. 
In terms of the force-freeness at the photosphere, that of NOAA 10969 has
$|F_z|/F_p>0.1$, while that of NOAA 11861 satisfies $|F_x|/F_p<0.1$,
$|F_y|/F_p<0.1$, and $|F_z|/F_p<0.1$.

 \begin{table}[H]
\centering
\caption{Force-freeness of the active regions from Equations (\ref{fx}), (\ref{fy}), and (\ref{fz})}
 \begin{tabular}{ccccc} 
\toprule
  & NOAA 10969  & NOAA 11861 \\
\toprule
 $|F_x|/F_p$& 0.018 &0.024\\
  $|F_y|/F_p$& 0.038 &0.071\\
   $|F_z|/F_p$& 0.43 &0.03 \\
\hline
\bottomrule
\end{tabular}
\label{sec3:table:forcefreeness}
\end{table}

\subsection{Vector Magnetic Fields in the Photosphere and the Chromosphere
\label{sec3:photo_chrom}}
\begin{figure}
\includegraphics[width=\columnwidth,clip,bb=0 0 1024 768]{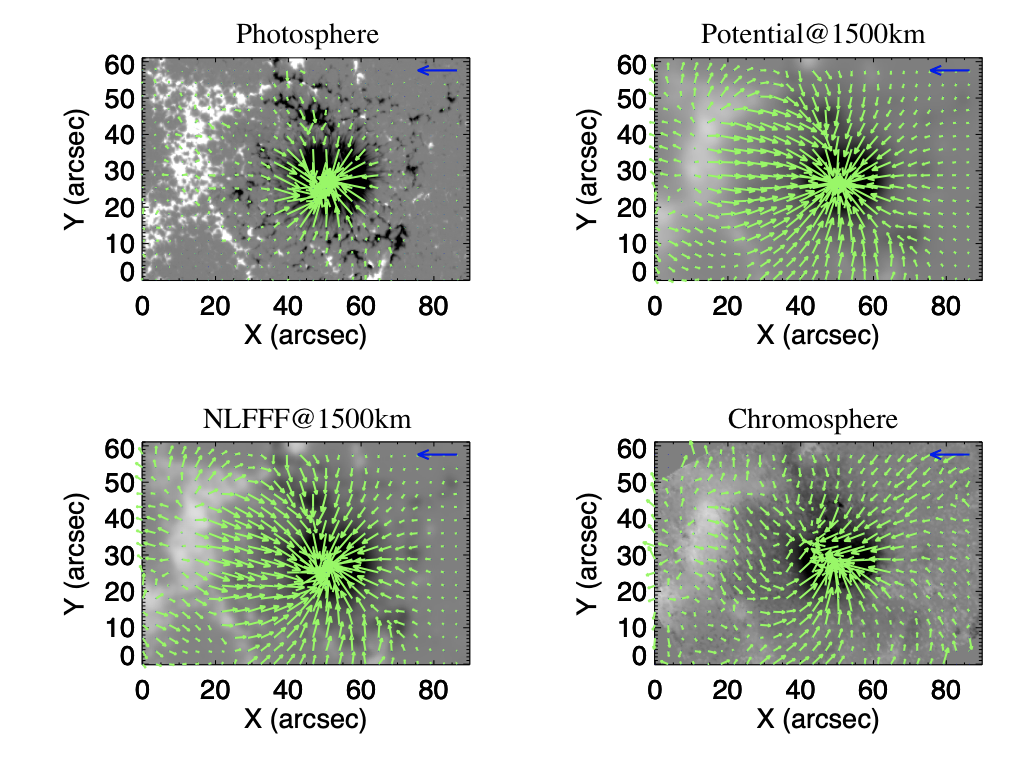}
\caption{ The vector magnetic field map of NOAA 10969. The gray scale shows the vertical magnetic field and the green arrows show the horizontal magnetic field. The length of blue arrow shows the field strength of 1500 G. Upper Left: Photospheric magnetic field observed with {\it Hinode}/SOT SP. Upper right: Potential field at 1500 km height. Lower left: NLFFF at 1500 km height. Lower right: Chromospheric magnetic field observed with VTT/TIP-2.}
\label{sec3:bxbybz_noaa10969}
\end{figure}

The upper left and lower right panels of Figure \ref{sec3:bxbybz_noaa10969}
show vector magnetic field maps at the photospheric and chromospheric heights
in NOAA 10969, respectively. The photospheric magnetic field is obtained from 
{\it Hinode} SOT/SP and the chromospheric field is derived from He {\sc i} 10830
\AA \ observation with the VTT/TIP-2. The background grayscale image shows the vertical
component of the magnetic field and the green arrows show the horizontal component of the magnetic field in the local frame. 
The horizontal magnetic field is displayed with a binning of $12\times12$ pixels. 

We only consider those pixels for which the linear polarization signals are
higher than $0.12\%$, to avoid any bias introduced by the noise.
Additionally, we only consider those pixels for which the inversion model
fulfills
\begin{equation}
\sigma_{\rm QU}=\frac{\frac{1}{n}\sum_{i=1}^{n} \sqrt{(Q_{\rm obs}-Q_{\rm syn})^2+(U_{\rm obs}-U_{\rm syn})^2}}{\sqrt{Q_{\rm peak}^2+U_{\rm peak}^2}}<0.08,
\label{qufit}
\end{equation}
where $n$ is the number of wavelength points considered during the HAZEL inversion, 
$Q_{\rm obs}$ and $U_{\rm obs}$ are the observed
Stokes profiles, $Q_{\rm syn}$ and $U_{\rm syn}$ are the synthetic Stokes
profiles, $Q_{\rm peak}$ and $U_{\rm peak}$ are the peak values of the linear
polarization. The pixels for which the linear polarization signals are small 
($<0.12\%$) or the fitting is poor ($\sigma_{\rm QU}> 0.08$), are treated as missing data. 
At photospheric height, strong horizontal magnetic fields are located in the negative
sunspot. The horizontal magnetic field in the sunspot is almost radial emerging from the
center of the sunspot. On the other hand, the horizontal magnetic field is
uniformly distributed at chromospheric heights and a twisted structure in a counterclockwise direction between positive and negative polarities can be seen.

\begin{figure}
\begin{center}
\includegraphics[width=0.9\columnwidth,clip,bb=0 0 592 768]{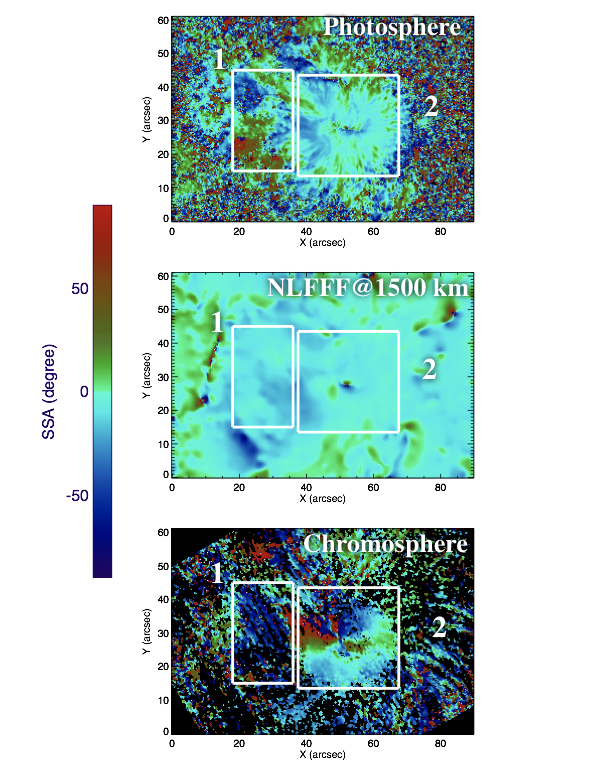}
\end{center}
\caption{The spatial distribution of signed shear angle (SSA) for NOAA 10969. From the top to the bottom for the photosphere, NLFFF at 1500km, chromosphere (He {\sc i} 10830 \AA). The regions where the LP signal is weak and the inversion did not fit the profiles well were masked by black color in the bottom panels. }
\label{sec3:ssa_noaa10969}
\end{figure}

\begin{figure}
\includegraphics[width=\columnwidth,clip,bb=0 0 1024 514]{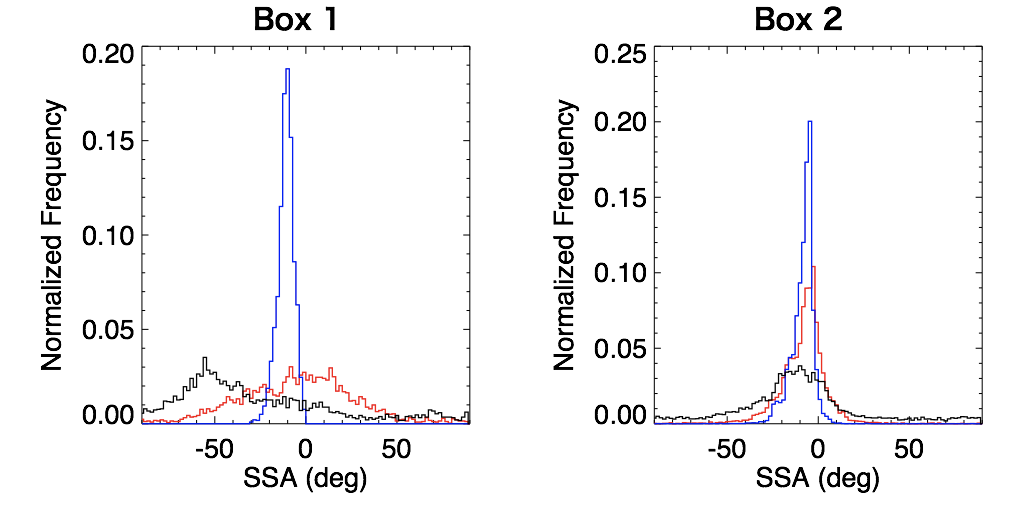}
\caption{The histograms of the SSA in the box 1 and box 2 in Figure \ref{sec3:ssa_noaa10969}. The black, red, and blue solid lines show the SSA from the chromospheric magnetic field derived from He {\sc i} 10830 \AA, the photospheric magnetic field, the NLFFF at the 1500 km height, respectively.}
\label{sec3:ssa_histo_noaa10969}
\end{figure}

Figure \ref{sec3:ssa_noaa10969} shows the spatial distribution of the SSA.
The top panel shows the SSA derived from Fe {\sc i} 6302 \AA \ observations with {\it Hinode}/SOT SP, whereas the bottom panel shows the SSA derived from He {\sc i} 10830 \AA \ observations with VTT. 
At chromospheric heights, the pixels where the linear polarization signal is weak
and the inversion did not fit the profiles well were masked and displayed in black colors. 
We focus on two regions (boxes 1 \& 2), where the chromospheric magnetic field is accurately derived. Box 1 
represents the region around the polarity inversion line and box 2 is above the 
sunspot with negative polarity. Figure \ref{sec3:ssa_histo_noaa10969} shows the
histograms of the SSA in both boxes. The black and red solid lines show the SSA from the
chromospheric magnetic field derived from He {\sc i} 10830 \AA \ and the
photospheric magnetic field, respectively. A clear deviation from potentiality is 
detected in box 1. While the peak value of the SSAs in the chromosphere in box 1 is around $-50^\circ$,
the SSAs at photospheric heights peak around $0^\circ$ although
with a very broad distribution when compared to that of the NLFFF extrapolation. The SSAs both in the
chromosphere and in the photosphere peak around $0^\circ$ for box 2. However,
while the SSAs at chromospheric heights in box 2 display a broad distribution, those
in the photosphere are much more concentrated around the mean value. 

\begin{figure}
\includegraphics[width=\columnwidth,clip,bb=0 0 1024 670]{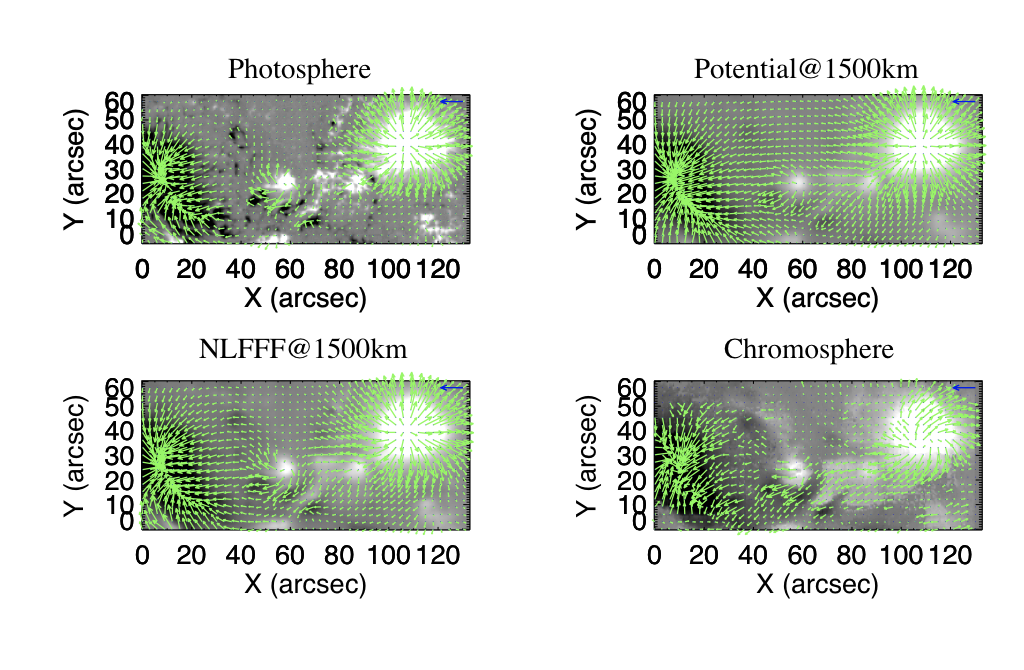}
\caption{Vector magnetic field distributions in NOAA 11861. The gray scale shows the vertical magnetic field and the green arrows show the horizontal magnetic field. The length of blue arrow shows the field strength of 1500 G. Upper Left: Photospheric magnetic field observed with {\it SDO}/HMI. Upper right: Potential field at 1500 km height. Lower left: NLFFF at 1500 km height. Lower right: Chromospheric magnetic field observed with DST/FIRS.}
\label{sec3:bxbybz_noaa11861}
\end{figure}

\begin{figure}
\begin{center}
\includegraphics[width=\columnwidth,clip,bb=0 0 664 768]{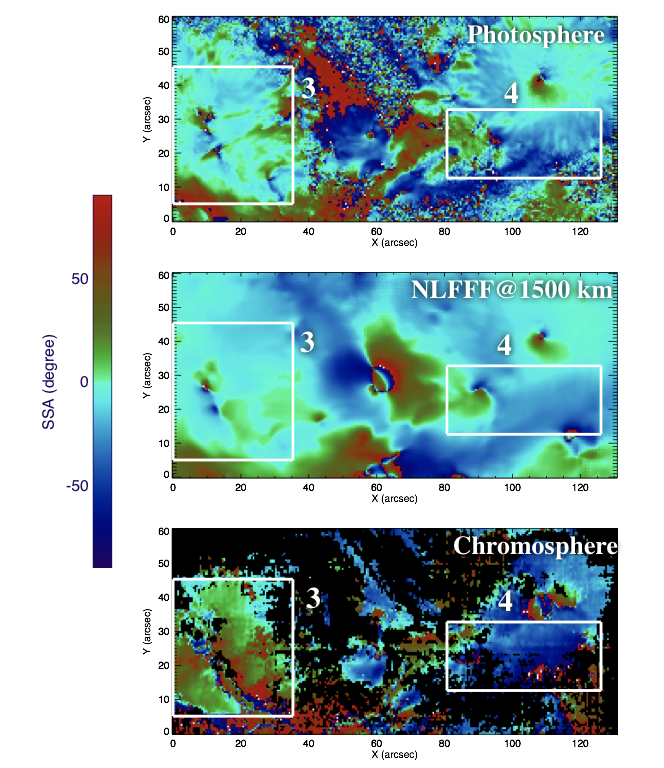}
\end{center}
\caption{The spatial distribution of signed shear angle (SSA) for NOAA 11861. From the top to the bottom for the photosphere, NLFFF at 1500km, chromosphere (He {\sc i} 10830 \AA). The regions where the LP signal is weak and the inversion did not fit the profiles well were masked by black color in the bottom panels. }
\label{sec3:ssa_noaa11861}
\end{figure}

\begin{figure}
\includegraphics[width=\columnwidth,clip,bb=0 0 1024 514]{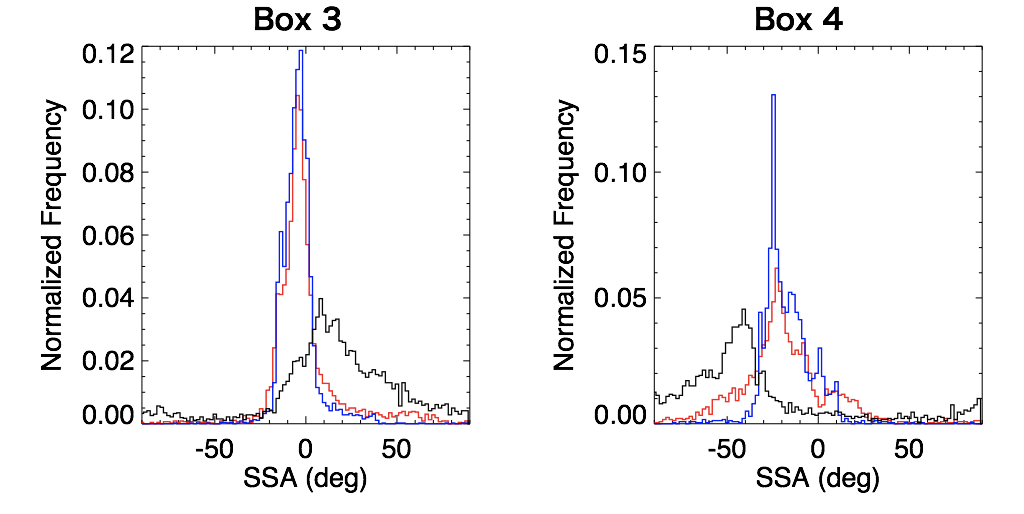}
\caption{The histograms of the SSA in the box 3 and box 4 in Figure \ref{sec3:ssa_noaa11861}. The black, red, and blue solid lines show the SSA from the chromospheric magnetic field derived from He {\sc i} 10830 \AA, the photospheric magnetic field, the NLFFF at the 1500 km height, respectively.}
\label{sec3:ssa_histo_noaa11861}
\end{figure}

Concerning NOAA 11861, the chromospheric magnetic field also shows larger non-potentiality than the photosphere.
Figure \ref{sec3:bxbybz_noaa11861} shows the spatial distribution of the vector magnetic field. 
The upper left and lower right panels show the photospheric field observed with {\it SDO}/HMI and the chromospheric field derived from the He {\sc i} 10830 \AA \ observation with the DST/FIRS, respectively.
The results of NOAA 10969 share some similarities with those of NOAA 11861. 
The horizontal magnetic field in the chromosphere looks more twisted compared to that of the photosphere in both the negative and positive sunspots. 
The difference between photospheric and chromospheric non-potentiality can be seen in Figure \ref{sec3:ssa_noaa11861}. 
The top and bottom panels show the SSA at the photospheric and chromospheric heights, respectively. 
The boxes 3 and 4 are located in the leading and following sunspots, respectively.
Figure \ref{sec3:ssa_histo_noaa11861} shows the histogram of the SSA in both boxes. 
In box 3 the SSAs in the chromosphere have mainly positive values while that of the photosphere peaks
around $0^\circ$. In box 4, the photosphere displays SSAs peaking around $-20^\circ$
while this increases in absolute value to $-40^\circ$ for the chromosphere. 

\subsection{Comparison with NLFFF at chromospheric heights
\label{sec3:chrom_nlfff}}
\begin{figure}
\includegraphics[width=\columnwidth,clip,bb=0 0 1417 566]{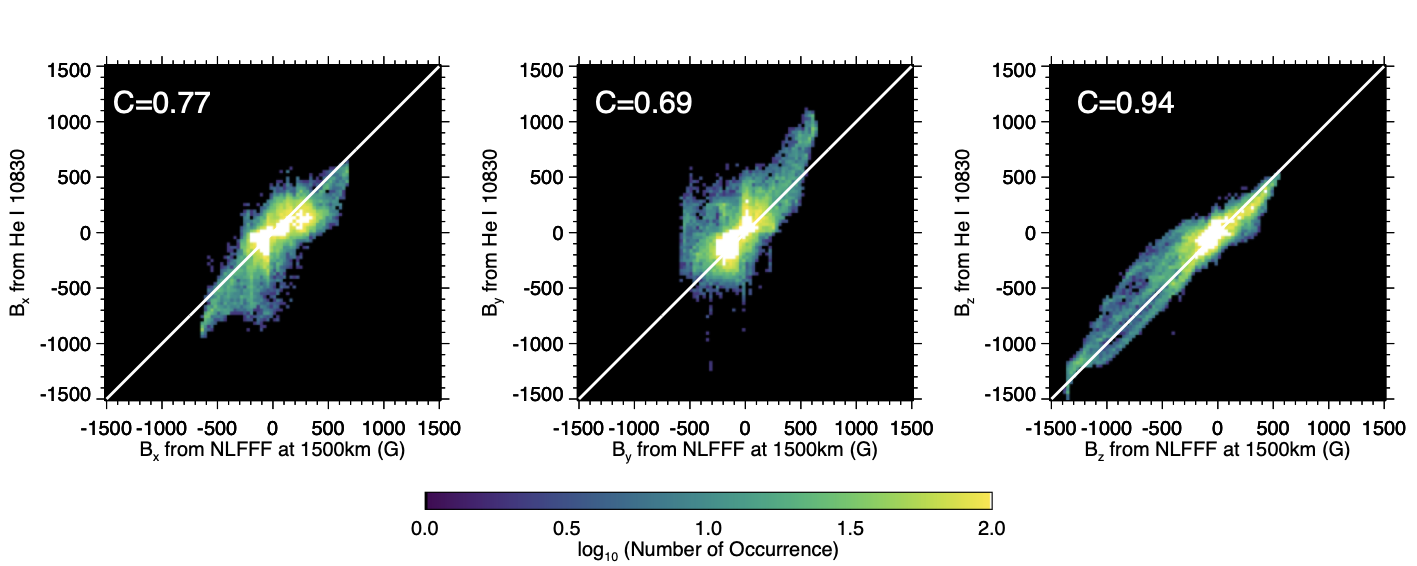}
\caption{The number density plots of the relation in $B_x$, $B_y$, and $B_z$ between NLFFF at 1500 km height and chromospheric magnetic field derived from He {\sc i} 10830 \AA \ in NOAA 10969.  This density plot is created with all pixels ($\sigma_{\rm qu}<0.08$).}
\label{sec3:bxbybz_hist_noaa10969}
\end{figure}

We compare the field obtained with the NLFFF extrapolation with that
inferred from the spectropolarimetric observations.
For the case of NOAA 10969, the upper right and lower left panels in Figure
\ref{sec3:bxbybz_noaa10969} show the potential field at 1500 km above the
photosphere derived from the observed photospheric field and the NLFFF,
respectively. The NLFFF at 1500km height displays 
horizontal magnetic fields very close to that of the potential extrapolation, pointing
to a small degree of non-potentiality at photospheric heights. 
However, the chromospheric magnetic field derived from the
inversion of the Stokes profiles on the He {\sc i} 10830 \AA\ multiplet 
shows a clear non-potential magnetic field vector
especially in the region between the positive and negative polarities. 
Figure \ref{sec3:bxbybz_hist_noaa10969} shows the joint probability distribution function (JPDF) between the cartesian components of the magnetic field ($B_x$, $B_y$, and $B_z$) from NLFFF at 1500 km above the photosphere and
that inferred from the spectropolarimetric observations. 
A good correlation in $B_z$ is present, with a Pearson correlation coefficient of
$C=0.94$. The absolute values of $B_z$ derived from the He {\sc i} data are
slightly smaller than those inferred from the NLFFF extrapolation. On the other hand, the horizontal
magnetic field ($B_x$ and $B_y$) shows comparatively weaker correlations,
$C=0.77$ and $0.69$, respectively.  We find larger (negative) values of $B_x$ inferred from 
He {\sc i} 10830 \AA \ when $B_x < -500$ G than those obtained from the extrapolation. 
Similarly, we also find larger (positive) values of $B_y$ inferred from 
He {\sc i} 10830 \AA \  when $B_y > 500$ G than those obtained from the extrapolation.
Strong horizontal magnetic fields, i.e., $(B_x^2+B_y^2)^{1/2}>500$ G, are located at the outer
part of the sunspot, so our results suggest that these regions have
stronger horizontal magnetic fields than those derived from the NLFFF
modeling. 

 When we compare only small SSA pixels ($|$SSA$|$ $<$ 5 degree in chromospheric magnetic field from He {\sc i} 10830 \AA), the correlation coefficient of horizontal magnetic field tends to increase, $C=0.86$ for $B_x$ and $C=0.77$ for $B_y$. 
 This is because the results of the NLFFF do not overestimate the SSA. 
 The field strength of horizontal and vertical field is not necessarily the same in some pixels. 
 Because the field strength has a dependency on the comparison height, the correlation coefficient also depends on the comparison height.

 In terms of the SSA, the NLFFF is clearly more potential than that derived from the He {\sc i} 10830 \AA\ observations as shown in the middle and bottom panels of Figure \ref{sec3:ssa_noaa10969}. 
This is also very clear from the blue histograms displayed in Figure
\ref{sec3:ssa_histo_noaa10969}. They peak at $-10^\circ$ for box 1 and
close to $0^\circ$ for box 2, both of them being much more narrow.

\begin{figure}
\includegraphics[width=\columnwidth,clip,bb=0 0 1417 566]{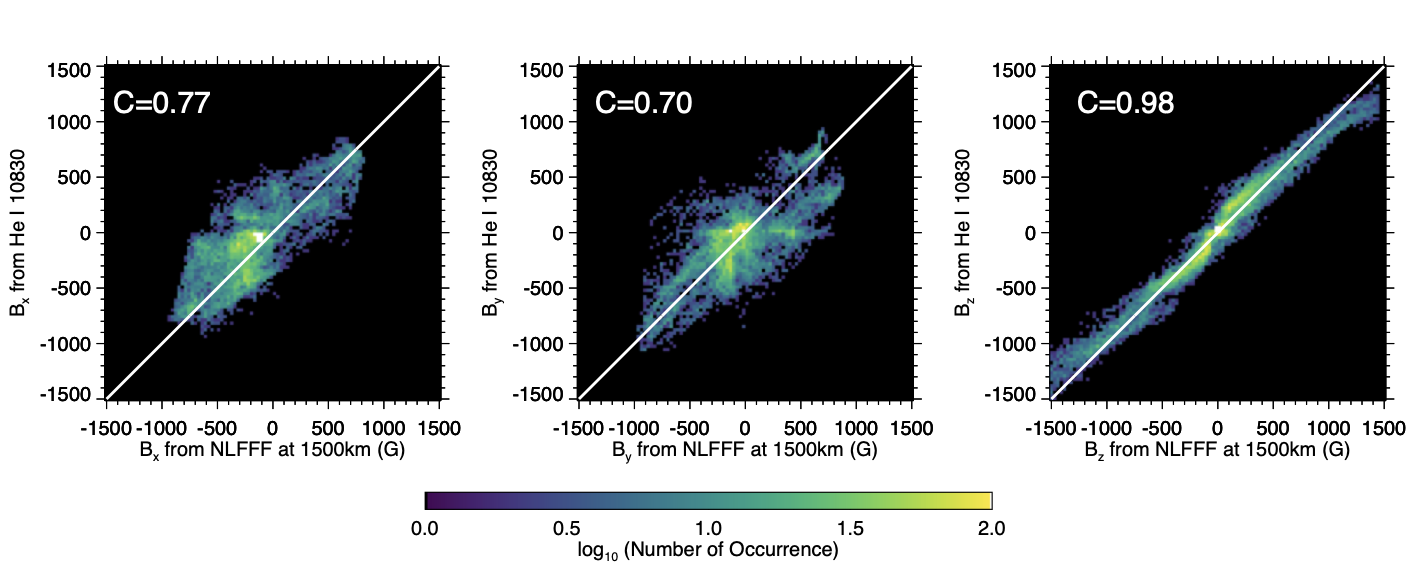}
\caption{The number density distribution in $B_x$, $B_y$, and $B_z$ of the relation between NLFFF at 1500 km height and chromospheric magnetic field derived from He {\sc i} 10830 \AA \ in NOAA 11861.  This density plot is created with  all pixels ($\sigma_{\rm qu}<0.08$).}
\label{sec3:bxbybz_hist_noaa11861}
\end{figure}

Concerning NOAA 11861, the upper right panel and lower left panel of Figure
\ref{sec3:bxbybz_noaa11861} show the potential field and the NLFFF extrapolation
at the 1500 km above the photosphere. While both show similar horizontal
magnetic fields, small deviations can be identified around
the center of the region of interest (ROI), $(x,y)\sim(50,20)$. A comparison with the 
chromospheric field inferred from He {\sc i} 10830 \AA\ yields clear
differences. 
The positive leading spot in the west side of the FOV in the box 3 shows a clear clockwise twist in the horizontal field from He {\sc i} 10830 \AA \ that
is absent from the NLFFF extrapolation. Figure
\ref{sec3:bxbybz_hist_noaa11861} shows the JPDFs for the cartesian
components of the magnetic field. Similar to
NOAA 10969, $B_z$ displays a tight correlation ($C=0.98$), while the horizontal components
show weaker correlations ($C=0.76$ for $B_x$ and $0.70$ for $B_y$). 
The strong vertical magnetic fields ($|B_z|>1000$ G) appear even
stronger in the NLFFF case. There is apparently no systematic bias in the 
horizontal components of the field, probably a consequence
of the much larger dispersion.

The histogram of the SSAs of the NLFFF is shown in blue solid in Figure 
\ref{sec3:ssa_histo_noaa11861}. The histogram of the NLFFF at 1500 km
is similar to that of the photospheric magnetic field. The
SSAs in the chromosphere have mainly positive values in box 3 while that of NLFFF is
around $0^\circ$ or negative. For box 4, the NLFFF extrapolation 
peaks at $-20^\circ$, while the one inferred from observations peaks around $-40^\circ$.

In summary, our analysis clearly indicates that the NLFFF extrapolation severely
underestimates the non-potentiality of the magnetic field at chromospheric heights.

\subsection{Coronal Loop Structures in Comparison with the Extrapolated Fields
\label{subsec:euv_nlfff}}
\begin{figure}
\includegraphics[width=\columnwidth,clip,bb=0 0 903 647]{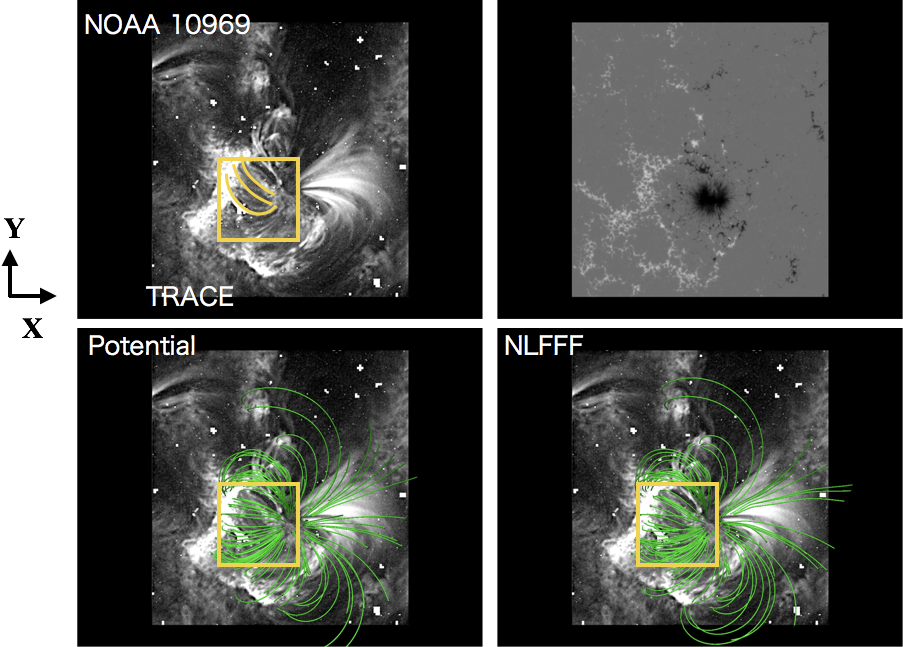}
\caption{The coronal magnetic field lines in NOAA 10969 extracted from an EUV 171 \AA \ image by visual inspection (upper left),  compared with the potential field lines (lower left) and NLFFF (lower left) based on the photospheric magnetic field (upper right). }
\label{sec3:euv_nlfff_noaa10969}
\end{figure}

Figure \ref{sec3:euv_nlfff_noaa10969} shows the qualitative comparison of
coronal field lines in NOAA 10969. The upper left panel shows  an EUV image from {\it TRACE} at 171 \AA. 
The yellow lines in the panel delineate the loops (arguably magnetic field lines) manually extracted by visual inspection. 
The upper right panel shows the vertical magnetic field in the photosphere obtained
with {\it Hinode} SOT/SP with the same FOV. The green solid lines overlaid on
the {\it TRACE} image in the lower panels show field lines estimated from
the potential field (bottom left) and NLFFF (bottom right) extrapolations. The field lines are
randomly selected in the computation box. The field lines of both extrapolations
show similar morphologies. However, there is a clear deviation with those that
we trace on the EUV image. 

\begin{figure}
\includegraphics[width=\columnwidth,clip,bb=0 0 903 647]{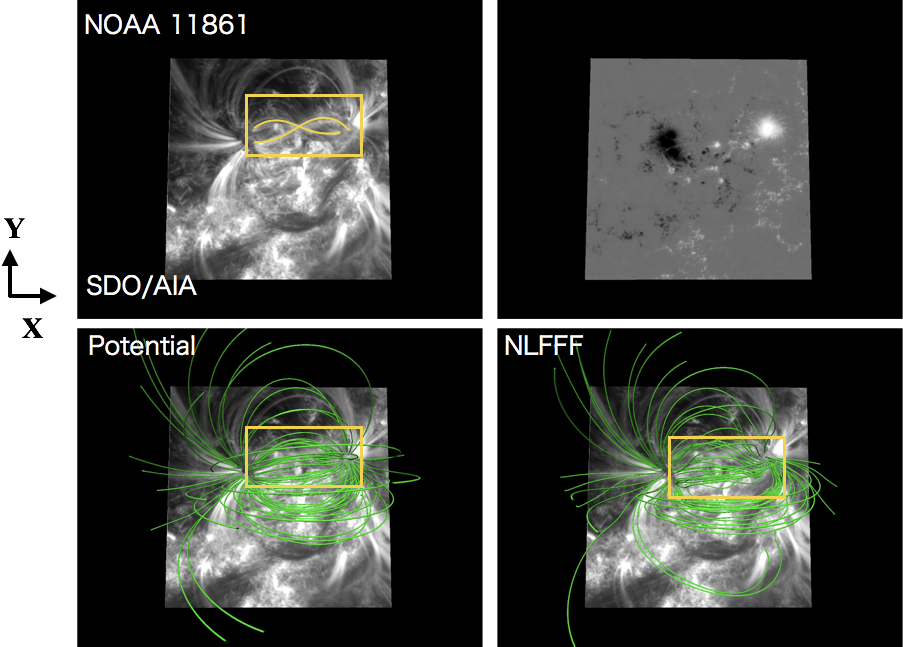}
\caption{Similar to Figure \ref{sec3:euv_nlfff_noaa10969}, but for NOAA 11861.}
\label{sec3:euv_nlfff_noaa11861}
\end{figure}

Concerning NOAA 11861, Figure \ref{sec3:euv_nlfff_noaa11861} shows the qualitative comparison.
Unlike the case of NOAA 10969, there is a clear difference between the potential and NLFFF 
extrapolations in the yellow box. The NLFFF field lines display a twisted structure that
is absent in the potential extrapolation. A somehow similar twist is found in the 
EUV image. As a consequence, the NLFFF qualitatively reproduces the 3D structure of the
magnetic field in this case.

\subsection{Relation between Chromospheric Vector Magnetic Fields and Fibril
Structures
\label{sec3:fibril}}
\begin{figure}
\includegraphics[width=\columnwidth,clip,bb=0 0 704 753]{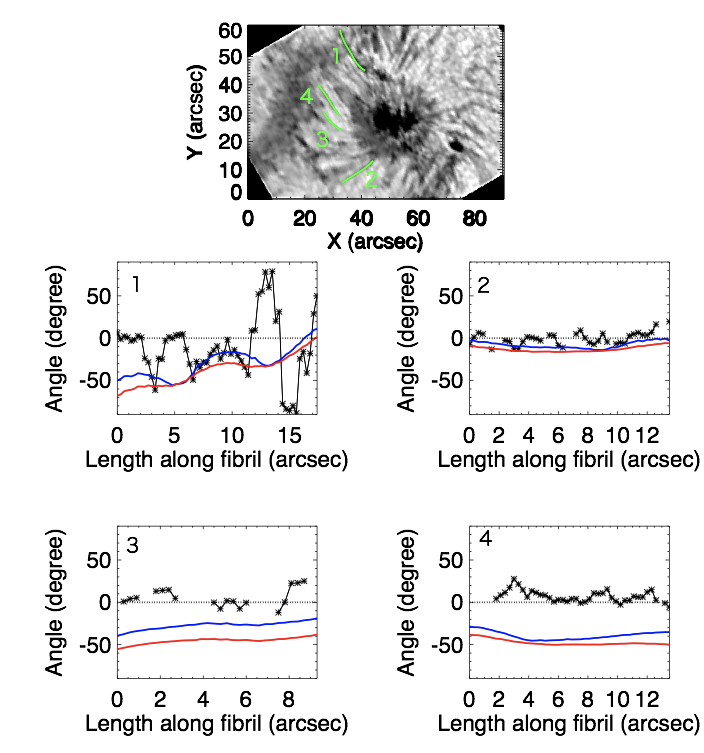}
\caption{The comparison between fibrillar structures and magnetic field vector. Upper panel:  The line core image of He {\sc i} 10830 \AA. Four green
lines show the  fibril structures automatically detected by OCCULT-2 code
\citep{2013Entrp..15.3007A}. Lower four panels: The angle between the fibril
structures and magnetic field vector along the fibril structures. Black lines
with asterisks, blue, and red lines show angles made by fibril structures with
the chromospheric magnetic field (He {\sc i} 10830\AA), the NLFFF at 1500 km,
and the potential field at 1500 km, respectively.}
\label{sec3:fibril_alignment}
\end{figure}

It is important to validate the results of the inferred chromospheric magnetic field with the
fibrillar structure seen in monochromatic images at the core of He {\sc i} 10830 \AA\ multiplet.
These structures, that can be seen in the upper middle panel of Figure \ref{sec3:observation}, are
dark fibrils found around the sunspot. It is often assumed that the magnetic field 
is aligned with the fibril structures.
Although theoretical and observational studies suggest that there is a misalignment 
due to partially ionization effects \citep{2011A&A...527L...8D, 2016ApJ...831L...1M,2017A&A...599A.133A},  
\cite{2013ApJ...768..111S} shows that the fibril structures are often well aligned with magnetic
field. 
Figure \ref{sec3:fibril_alignment} shows the comparison between the inferred 
magnetic field and the chromospheric features seen in the core of He \textsc{i} 10830 \AA\
for regions inside box 1. The image in the upper panel shows the intensity at the line core of
He {\sc i} 10830 \AA. The four green lines show the fibrils automatically
detected by the OCCULT-2 code \citep{2013Entrp..15.3007A}. The lower four panels
show the angle between the fibrillar structure and the magnetic field vector along each
fibrillar structure. In each plot we display the angle between the fibril and
the inferred magnetic field (in black lines with symbols), the NLFFF extrapolation (blue lines)
and the potential extrapolation (red lines) at 1500 km. Except for fibril 1, the magnetic field vectors 
derived from He {\sc i} 10830 \AA \ are very well aligned with the fibrils. The misalignment
between the fibril and the extrapolations can easily reach $50^\circ$.

\subsection{Height dependence
\label{sec3:evaluation}}
\begin{figure}
\includegraphics[width=\columnwidth,clip,bb=0 0 817 743]{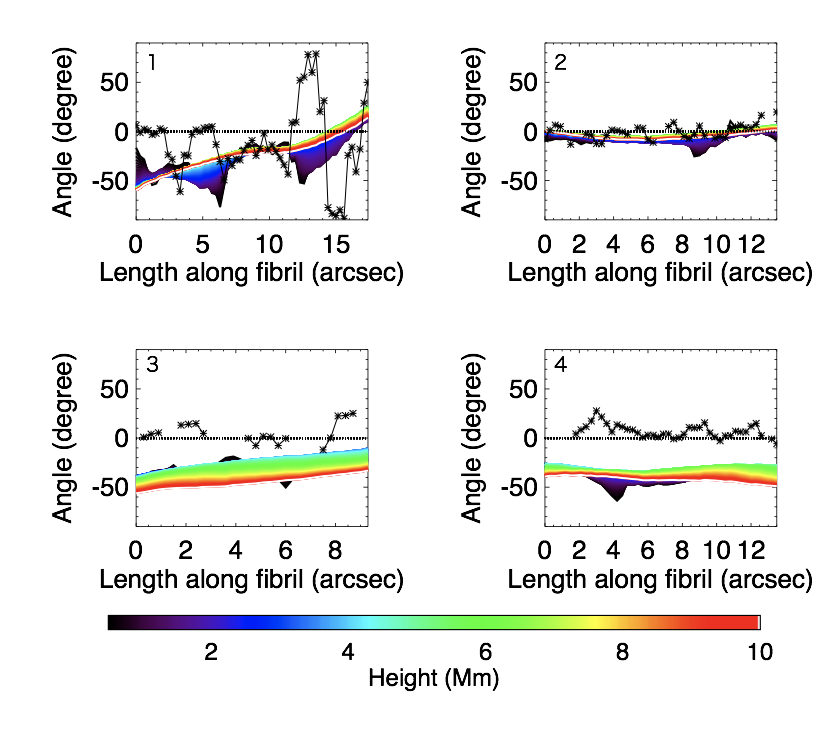}
\caption{Height dependence of magnetic field azimuth. Similar to Figure \ref{sec3:fibril_alignment}, each panel shows the angle between magnetic field vector and fibril structures. Color shows the angle between NLFFF at the height of 0-10 Mm and fibril structures. Black solid lines with starts show the angle between magnetic field vector from He {\sc i} 10830 \AA \ and the fibril structures.}
\label{sec3:fibril_alignment_all_height}
\end{figure}

Although the formation layer of He {\sc i} 10830 \AA \ is considered to be thin in
chromospheric structures, the formation height of He {\sc i} 10830 \AA \ may vary by a
large margin (hundreds or even thousands of km) depending on the locations of the active
regions. 
Although we have used all extrapolations at 1500 km, there is a possibility that 
a fair comparison would require to use different heights. 
To test this, Figure \ref{sec3:fibril_alignment_all_height} shows what happens 
when the extrapolations are computed at different heights between 0 and 10 Mm. These
results show that the difference between the extrapolated and measured magnetic
fields remains even when different heights are considered.

\section{Discussions
 \label{sec3:discussion}}
In this paper, the vector magnetic field observations at both photospheric and chromospheric
heights suggest that the 
chromospheric magnetic field may have larger non-potentiality than the photospheric
magnetic field. \cite{2017A&A...604A..98J} also investigated photospheric and
chromospheric magnetic fields of two simple round sunspots by using the Si {\sc i} 10827 \AA \
and He {\sc i} 10830 \AA\ lines and suggested the possibility that the
chromospheric magnetic field has a larger twist compared to the photospheric magnetic
field. Our study has extended their view by examining the entire active
regions, not restricted to a simple sunspot. Large FOV observations allow us to
identify twisted structures more clearly as shown in Figures
\ref{sec3:bxbybz_noaa10969} and \ref{sec3:bxbybz_noaa11861}.
\cite{2012ApJ...748...23Y} also extrapolated the 3D magnetic field from both
the photosphere and chromosphere. While they performed qualitative comparison
of the 3D structure of the field lines, we have quantitatively compared the
non-potentiality of the magnetic field measured by the SSA.
Compared with the measurements of the chromospheric magnetic field, we revealed
that the NLFFF modeling  may underestimate the non-potentiality both in active
regions NOAA 10969 and 11861.
We have to note that the NLFFF extrapolation may depend on the method used even with the same photospheric boundary condition \citep{2009ApJ...696.1780D}.  
 The possibility of the extrapolation method dependency can not be rejected in this study and remained for future works.

It is mandatory to discuss the possible influence of the azimuth ambiguities in
our study. Apart from the well-known 180$^\circ$ ambiguity, which is also
present in the transverse Zeeman effect 
\citep{2004ASSL..307.....L}, a second type of ambiguity, termed Van Vleck ambiguity
\citep{2004ASSL..307.....L, 2008ApJ...683..542A} appears. The Van Vleck ambiguity occurs
only as a consequence of the Hanle effect. In this case, potentially up to four (at $\pm 90^\circ$ and $\pm 180^\circ$)
possible azimuths in the plane of the sky can lead to the same polarimetric signal. 
Obviously, the Van Vleck ambiguity disappears when the linear polarization signals are dominated 
by the transverse Zeeman effect, such as in the edge of sunspots. 
On the contrary, the Van Vleck ambiguity may exist in the other regions.
We checked the validity of the azimuth for such pixels by comparing with the fibril structures as shown in Figure \ref{sec3:fibril_alignment}.

\begin{figure}
\includegraphics[width=\columnwidth,clip,bb=0 0 414 356]{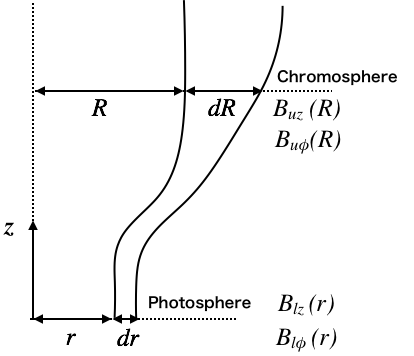}
\caption{Sketch of the expansion of the magnetic flux tube. Solid lines show the magnetic field lines. }
\label{ponchi_parker}
\end{figure}

There are two possibilities to cause the underestimation of the
non-potentiality when using NLFFF modeling. The first cause is the vertical gas
pressure gradient in the lower atmosphere, which is related to the non-force-freeness in the photosphere. \cite{1974ApJ...191..245P}
investigated the radial expansion of a magnetic flux tube due to the decrease of
the gas pressure with height, as shown in Figure \ref{ponchi_parker}.
The conservation of the longitudinal magnetic flux gives
\begin{equation}
B_{lz}rdr=B_{uz}RdR,
\end{equation}
where $B_{lz}$ and $B_{uz}$ are the longitudinal magnetic fields at the lower
and upper atmospheres, respectively, and $r$ and $R$ are the radial distances
from the axis at the lower and upper atmospheres, respectively. The conservation
of the torque of the azimuthal Maxwell stress yields
\begin{equation}
r(B_{l\phi}B_{lz})rdr=R(B_{u\phi}B_{uz})RdR,
\end{equation}
where $B_{l\phi}$ and $B_{u\phi}$ are the azimuthal magnetic fields at the lower
and upper atmospheres, respectively. Combining these two equations, one can calculate
the ratio between $T_{u}$ and $T_{l}$, the number of turns per unit length for torsional
equilibrium at the upper and lower atmospheres, respectively:
\begin{eqnarray}
\frac{T_{u}}{T_{l}}&=&\frac{B_{u\phi}/(2\pi RB_{uz})}{B_{l\phi}/(2\pi r
B_{lz})}, \nonumber \\
&=&\frac{r}{R}\frac{dR}{dr}.
\end{eqnarray}
Assuming that the expansion rate of the flux tube becomes larger at the large radius,
$R/r<dR/dr$, the number of turns per unit length becomes larger at the upper
atmosphere, $T_u/T_l>1$. This means that the expansion of the flux tube
enhances the non-potentiality at chromospheric heights and suggests that
the gas pressure significantly affects the magnetic field even in active
regions. Since NLFFF calculation is based on the photospheric magnetic field in
this study, the effect of expansion of the magnetic flux tubes can not be
reproduced. 
 The underestimation can be found even in NOAA 11861, which satisfies force-free condition evaluated with Equations (4), (5), and (6).
We have to note that the force-free conditions in Equations (4), (5), and (6) are not sufficient conditions but necessary conditions.
Our results imply that even though the the force-free conditions in Equations (4), (5), and (6) are satisfied, the effect of force-freeness in the photosphere can be found in the upper atmosphere.

 The second possible cause is the uncertainty in the magnetic field observations of the
photospheric layer in the penumbral regions. 
\cite{2002ApJ...575.1131L} reported that the magnetic field in the penumbral
region has fluted structure.
In other words, the penumbra may have two magnetic field components in one pixel.
We inverted the photospheric lines Fe {\sc i} 6301.5 \AA \ and 6302.5 \AA \ with
only a single Milne-Eddington component. 
The presence of several components on the same pixel might lead to a biased estimation of the azimuth of the magnetic field in the penumbral regions.

Our NLFFF modeling and its comparison with the inferred magnetic field strongly suggest that the non-potentiality in active regions may be larger than that previously estimated based on the NLFFF extrapolation not only in the chromosphere but also in the corona. 
A quantitative estimation of the non-potentiality in the upper atmosphere is important in understanding the onset mechanism for solar flares.
Magneto-hydrodynamical instabilities are considered to be important mechanisms for the onset of the eruption of magnetic flux rope and are sensitive to the 3D magnetic field structure. 
As one of such instabilities, the kink instability \citep{2004A&A...413L..27T} occurs when the twist of the magnetic
flux rope exceeds a critical value, so that it can be characterized by
the twist number $T_{w}$ \citep{2006JPhA...39.8321B} of the flux rope,
\begin{equation}
T_w=\frac{1}{4\pi}\int\alpha dl,
\label{eqn:twistnumber}
\end{equation}
where $l$ is the length along the magnetic field line and $\alpha$ is the force-free parameter. 
 \cite{2018ApJ...864..138J} performed a statistical study of the relation among
the CME occurrence and the presence of MHD instabilities (torus and kink instabilities) 
based on NLFFF extrapolation. Their conclusion is that kink
instabilities play little role in discriminating between confined and ejective
events.


Our results show that the magnitude of the underestimation of the
non-potentiality is different from each other in the two active regions, as
shown in Figures \ref{sec3:ssa_histo_noaa10969} and
\ref{sec3:ssa_histo_noaa11861}. In the linear force free
case (LFF), it is relatively easy to compute the twist number
as a function of the SSA. To this end, the magnetic field 
in the LFF case is given by:
\begin{equation}
\vector{B}=\left(
\begin{array}{c}
B_x\\
B_y\\
B_z
\end{array}
\right)=B_0\left(
\begin{array}{c}
\alpha_0 k^{-1} \cos ky\\
-l k^{-1} \cos ky \\
\sin ky
\end{array}
\right)\exp(-lz),
\end{equation}
where $\alpha_0$ is the constant force-free parameter, k is the wave number,
$l=(k^2-\alpha_0^2)^{1/2}$, and $B_0$ is a constant. As usual, when $\alpha_0=0$, the
magnetic field becomes potential field and points along 
the $y$-axis. In this case, the SSA can be
calculated as
\begin{eqnarray}
 {\rm SSA}&=&\tan^{-1} (-B_x/B_y),\\
 &=&\tan^{-1} (\alpha_0/l),
 \end{eqnarray}
so that
\begin{equation}
\alpha_0=\pm k \sin ({\rm SSA}).
\end{equation}
Because $\alpha_0$ is constant along
the field line, the twist number is given by
\begin{eqnarray}
T_w&\sim& \frac{1}{4\pi} \alpha L = \pm\frac{kL}{4\pi}\sin(\mathrm{SSA}),
\end{eqnarray}
where $L$ is the length of the field line. Therefore, when the SSA 
increases from $10^\circ$ to $40^\circ$, the twist number increases 
by a factor 3.7. Likewise, when the SSA
increases from $30^\circ$ to $50^\circ$, the twist number only increases by a
factor 1.5. If this behavior is approximately maintained in the NLFFF
case, we find an underestimation of a factor $\sim 2$ of the twist number
for NOAA 11158, which goes up to more than 3 for the case of 
NOAA 10969. 
This will significantly affect the probability of the occurrence of ejective flares in these regions.

  Although our results reveal the incompleteness in the current NLFFF modelings,  we do not conclude that the extrapolation method is unnecessary.  
 There are several points to discuss the topic. 
Firstly, our results are based on the analysis of two active regions. 
It is too early to conclude that the extrapolation method is not reliable. 
Further investigations with statistical analysis are required. 
Secondly, the main cause of the failure has a possibility to be solved. 
We attribute the failure of the current NLFFF modeling to the non-force-freeness (high plasma-beta) in the photosphere. 
The plasma beta decreases as the height increases and is thought to be sufficiently small at the chromospheric height \citep{2001SoPh..203...71G}. 
Therefore, the inclusion of chromospheric magnetic field to the NLFFF modeling could improve the NLFFF modeling. 
Thirdly, the reliable extrapolation results in the solar corona are not guaranteed even if we use the chromospheric magnetic field as another constraint. 
The extrapolation methods have other problems to be solved in addition to the non-force-freeness in the photosphere. 
The results of the NLFFF extrapolation have the dependency on the method \citep{2009ApJ...696.1780D}, spatial resolution \citep{2015ApJ...811..107D}, and initial guess \citep{2020arXiv200500177K}. More efforts should be made to solve such problems and develop more reliable extrapolation method. 
Fourthly, the coronal magnetic fields are expected to be measured by the Daniel K. Inouye Solar Telescope \citep[DKIST:][]{2011ASPC..437..319K}. 
Although DKIST will provide fascinating data of the coronal magnetic field, we think that the extrapolation method is still necessary. The extrapolation method and the direct coronal magnetic field measurements are complementary. While the coronal measurements of DKIST is limited to the limb, the extrapolation methods are usually applied to on-disk observations. Utilizing both information of the extrapolation and direct measurements is important to tackle the questions of the solar corona.

\section{Summary}
We examined the chromospheric magnetic fields from spectropolarimetric observations 
of the He {\sc i} 10830 \AA\ multiplet, which were compared with the 
chromospheric magnetic field extrapolated from the photospheric magnetic field. 
Our main conclusions are:
\begin{description}
\item[(1)] The chromospheric magnetic field derived from the spectropolarimetric observations shows more twisted magnetic fields at several locations than the photospheric magnetic field does.
\item[(2)] The potential field and NLFFF extrapolation from the photospheric magnetic field 
underestimate the non-potentiality at the chromospheric height at many locations.
\end{description}

From the analysis for two active regions, we have revealed that the magnetic 
field in the upper atmosphere may have higher non-potentiality than previously 
thought based on the NLFFF modeling.
Our studies emphasize the importance of the chromospheric magnetic field measurements 
for more accurate 3D magnetic field modeling and the understanding of the 
non-potentiality in active region corona.
Because the non-potentiality is a crucial ingredient in investigating the MHD instability, 
our findings are important for understanding the onset mechanisms for solar flares
and CMEs, which affect the environment in the solar system.
In the current state, the chromospheric magnetic field observations in active regions 
are very few in number.
It is obvious that we should make efforts to perform more observations
of the chromospheric magnetic fields in flare-productive active regions with 
future large aperture telescopes.

{\bf Acknowledgements}
The authors thank to the referee for useful comments.
{\it Hinode} is a Japanese mission developed and launched by ISAS/JAXA,
collaboratoing with NAOJ as a domestic partner, NASA and STFC (UK) as
international partners. It is operated by these agencies in cooperation with ESA
and NSC (Norway). We gratefully acknowledge the {\it SDO}/HMI team for providing
data. Our calculations of NLFFF modeling were performed on JAXA Supercomputer
System generation 2 (JSS2). This publication makes use of data obtained during
Cycle 2 DST Service Mode Operations under the proposal ID P495. This work was
supported by MEXT/JSPS KAKENHI Grant Numbers JP15H05814 and JP18H05234. AAR acknowledges financial 
support from the Spanish Ministerio de Ciencia, Innovaci\'on y Universidades 
through project PGC2018-102108-B-I00 and FEDER funds.

\end{document}